\documentclass[a4paper,english,showpacs,aps,superscriptaddress]{revtex4}
\usepackage[T1]{fontenc}
\usepackage{amsmath}
\usepackage[dvips]{graphicx}
\usepackage{amssymb}

\makeatletter
\usepackage{babel}
\makeatother
\begin{document}

\title{Energy transfers and locality in magnetohydrodynamic turbulence}

\author{Mahendra K. Verma}

\affiliation{Department of Physics, Indian Institute of Technology, Kanpur --
208016, INDIA}

\author{Arvind Ayyer}

\affiliation{Department of Physics and Astronomy, Rutgers University, Piscataway,
NJ, USA}

\author{Amar V. Chandra}

\affiliation{Department of Physics, Indian Institute of Technology, Kanpur --
208016, INDIA}

\date{March 7, 2005}

\begin{abstract}
The shell-to-shell energy transfer rates for magnetohydrodynamic (MHD)
turbulence are computed analytically, which shows local energy transfer
rates from velocity to velocity, velocity to magnetic, magnetic to
velocity, and magnetic to magnetic fields for nonhelical MHD in the
inertial range. It is also found that for kinetic-energy dominated
MHD fluid, there is a preferential shell-to-shell energy transfer
from kinetic to magnetic energy; the transfer is reversed for magnetic-energy
dominated MHD fluid. This property is the reason for the asymptotic
value of Alfvén ratio to be close to 0.5. The analytical results are
in close agreement with recent numerical results. When magnetic and
kinetic helicities are turned on, the helical contributions are opposite
to the corresponding nonhelical contributions. The helical energy
transfers have significant nonlocal components. 
\end{abstract}


\maketitle

\section{Introduction}

Turbulent fluid and plasma flows exhibit complex behaviour. One such
phenomena is the energy transfers among various scales. For fluid
turbulence the energy transfer issues have been investigated in great
details. However, detailed analysis of these processes is lacking
in magnetohydrodynamic (MHD) turbulence. Detailed understanding of
energy transfer is useful for understanding various physical process,
for example, dynamo mechanism to generate magnetic field in astrophysical
objects. These results are also useful in modelling MHD flows and
in simulations. For example, we need to model backscatter and forward
energy transfer for large-eddy simulations. In the present paper we
investigate the above issues analytically.

Kolmogorov's fluid turbulence phenomenology for incompressible turbulence
is based on local energy transfer between wavenumber shells. There
are several quantitative theories in fluid turbulence about the amount
of energy transfer between neighbouring wavenumber shells. For examples,
Kraichnan \cite{Krai:71} showed that $35$\% of the energy flux comes
from wavenumber triads where the smallest wave-number is greater than
one-half of the middle wavenumber. This phenomenology has been verified
using numerical and analytical methods \cite{Doma:Local2,Zhou:Local,Zhou:RevLocal,Ayye}.
Debliquy et al. \cite{Oliv:Simulation} recently studied the issues
of energy transfers in decaying magnetohydrodynamic (MHD) turbulence
using direct numerical simulation (DNS). Alexakis et al. \cite{Mini:I}
and Mininni et al. \cite{Mini:II} performed the similar calculations
for forced MHD turbulence for both helical and nonhelical flows. They
found that typically, the shell-to-shell energy transfer is local.
In the present paper we compute the above quantities analytically,
and compare them with the numerical results. 

The interactions in MHD are through $(\mathbf{u}(\mathbf{k}),\mathbf{u}(\mathbf{p}),\mathbf{u}(\mathbf{k-p}))$
and $(\mathbf{b}(\mathbf{k}),\mathbf{u}(\mathbf{p}),\mathbf{b}(\mathbf{k-p}))$
triads, where $\mathbf{(u},\mathbf{b})$ are the velocity and magnetic
fields respectively, and $\mathbf{k}$, $\mathbf{p}$, and $\mathbf{k-p}$
are the wavenumbers of the triad. Kraichnan \cite{Krai:71} gave a
general formalism to compute the magnitudes of triad interactions
using transfer function $S(k|p,q)$ \cite{Lesi:book}. In this paper
we will compute the shell-to-shell energy transfer using a modified
method called \emph{mode-to-mode energy transfer rate} $S(k|p|q)$,
which represents the energy transfer mode from \textbf{$\mathbf{p}$}
to mode $\mathbf{k}$, with mode \textbf{$\mathbf{q}$} acting as
a mediator. The new formalism is necessary for computing the shell-to-shell
energy transfer because the earlier formalism suffers from ambiguity
arising due to the third leg of the interaction (see \cite{Dar:flux,MKV:PR}).
The calculation is done using perturbative field-theory up to first-order
in perturbation. We take Kolmogorov's spectrum for the energy spectrum
as discussed in current numerical and analytical papers \cite{MKV:B0_RG,Srid1,Srid2,MKV:MHD_PRE,MKV:MHD_RG,Hnat}.
Note that the field-theoretic calculations has a lot of similarity
with Eddy-damped quasi-normal Markovian (EDQNM) approximation.

MHD turbulence involves interactions among velocity and magnetic modes,
hence energy transfer takes place between velocity to velocity, magnetic
to magnetic, velocity to magnetic, and magnetic to velocity modes.
Debliquy et al. \cite{Oliv:Simulation} computed the shell-to-shell
energy transfers in decaying MHD turbulence using simulation data
on $512^{3}$. In their calculation cross helicity, magnetic helicity,
and kinetic helicity are negligible. They also took the mean magnetic
field to zero. Debliquy et al. found forward and local energy transfer
from velocity to velocity, and magnetic to magnetic fields. Regarding
the velocity-to-magnetic energy transfer, for the Alfvén ratio greater
than approximately 0.4, the energy transfer is from kinetic to magnetic;
the transfer direction is reversed for Alfvén ratio less than 0.4.
Dar et al. \cite{Dar:flux}, Alexakis et al. \cite{Mini:I} and Mininni
et al. \cite{Mini:II} have done the similar analysis for 2D and 3D
forced MHD turbulence, with forcing at small wavenumber of velocity.
They find local energy transfer for velocity to velocity fields, and
magnetic to magnetic fields in the inertial range. However, the small-wavenumber
velocity shells provide energy to the small-wavenumber magnetic shells,
as well as to the inertial range magnetic shells. We will show in
this paper that our theoretical results on inertial range energy transfers
are in general agreement with the above numerical results.

Pouquet et al. \cite{Pouq:EDQNM} were the first to investigate whether
interactions in MHD turbulence are local or not. Their analysis is
based on eddy-damped quasi-normal Markovian (EDQNM) calculation. They
claimed that nonlocal interactions exist in MHD due to the mean magnetic
field (Alfv\'{e}n effect) and helicity. According to Pouquet et al.,
the local interactions cause the energy cascade, but the nonlocal
ones lead to an equipartition of kinetic and magnetic energy. In the
present paper we will also show that helicity induces nonlocal energy
transfers.

A detailed picture of energy transfers is very useful for understanding
turbulence and its modelling. In this paper we will show how we can
use our theoretical results to argue why the asymptotic state of MHD
turbulence is close to 0.5. The detailed shell-to-shell energy transfer
also provide us important ideas for large-eddy simulations (see Debliquy
et al. \cite{Oliv:Simulation} for connection with large-eddy simulations)
and EDQNM calculation, which assumes local energy transfers among
wavenumber shells. 

It is well known that compressible turbulence involves energy transfers
from pressure fluctuations to the velocity and magnetic field \cite{LandFlui:book,Zank:Compress_PoF,Zank:Compress_PRL,Cho:Compressible,MKV:Rev}.
The theoretical, numerical, and observational studies show that the
energy spectrum deviates from Kolmogorov's spectrum. For example,
Burgers equation, which represents fully compressible fluid, has energy
spectrum proportional to $k^{-2}$. The theory of compressible turbulence
is not yet developed as much as that for incompressible turbulence.
Due to the uncertainty of energy spectrum and other properties in
the inertial range, in this paper we have confined ourselves to the
study of shell-to-shell energy transfer for incompressible turbulence
only.

The outline of the paper is as follows: In Sec. 2 we compute the shell-to-shell
energy transfer rates for nonhelical and helical MHD. In Sec. 3 we
use our results to show why the asymptotic state of MHD turbulence
has Alfvén ratio close to 0.5. The last section, Sec. 4, contains
conclusions.

\section{Calculation of the Shell-to-shell Energy Transfers }

In MHD turbulence, velocity $\mathbf{(u)}$ and magnetic fields $\mathbf{b}$
interact with each other and among itself to produce complex energy
transfers. The energy exchange can take place between a $\mathbf{u}$
Fourier mode to $\mathbf{u}$ Fourier mode, between a $\mathbf{b}$
Fourier mode to $\mathbf{b}$ Fourier mode, or between a $\mathbf{u}$
Fourier mode to $\mathbf{b}$ Fourier mode. These transfers are studied
using Kraichnan's formula $S(\mathbf{k}|\mathbf{p},\mathbf{q})$ \cite{Lesi:book}
or \emph{mode-to-mode energy transfer rate} $S(\mathbf{k}|\mathbf{p,q})$.
In this paper we use the \emph{mode-to-mode energy transfer rates}
$(S^{YX}(\mathbf{k'}|\mathbf{p}|\mathbf{q}))$ that represents the
energy transfer rates from mode \textbf{$\mathbf{p}$} of field $X$
to mode $\mathbf{k}$ of field $Y$, with mode \textbf{$\mathbf{q}$}
acting as a mediator \cite{Dar:flux,MKV:PR}. Note that $\mathbf{k}'+\mathbf{p}+\mathbf{q}=0$.
This formalism is used for computing the shell-to-shell energy transfer
because the earlier formalism suffers from ambiguity arising due to
the third leg of the interaction (see \cite{Dar:flux,MKV:PR}). The
mode-to-mode energy transfer rates in MHD turbulence are given by\begin{eqnarray*}
S^{uu}(\mathbf{k'}|\mathbf{p}|\mathbf{q}) & = & -\Im\left(\left[\mathbf{k'}\cdot\mathbf{u}(\mathbf{q)}\right]\left[\mathbf{u}(\mathbf{k')}\cdot\mathbf{u}(\mathbf{p})\right]\right),\\
S^{bb}(\mathbf{k'}|\mathbf{p}|\mathbf{q}) & = & -\Im\left(\left[\mathbf{k'}\cdot\mathbf{u}(\mathbf{q)}\right]\left[\mathbf{b}(\mathbf{k')}\cdot\mathbf{b}(\mathbf{p})\right]\right),\\
S^{ub}(\mathbf{k'}|\mathbf{p}|\mathbf{q}) & = & \Im\left(\left[\mathbf{k'}\cdot\mathbf{b}(\mathbf{q)}\right]\left[\mathbf{u}(\mathbf{k')}\cdot\mathbf{b}(\mathbf{p})\right]\right),\\
S^{bu}(\mathbf{k'}|\mathbf{p}|\mathbf{q}) & = & \Im\left(\left[\mathbf{k'}\cdot\mathbf{b}(\mathbf{q)}\right]\left[\mathbf{b}(\mathbf{k')}\cdot\mathbf{u}(\mathbf{p})\right]\right),\end{eqnarray*}
where the above four formulas denote the energy transfers from $\mathbf{u}(\mathbf{p})$
to $\mathbf{u}(\mathbf{k})$, from $\mathbf{b}(\mathbf{p})$ to $\mathbf{b}(\mathbf{k})$,
from $\mathbf{b}(\mathbf{p})$ to $\mathbf{u}(\mathbf{k})$, and from
$\mathbf{u}(\mathbf{p})$ to $\mathbf{b}(\mathbf{k})$ respectively.
For the derivation the reader is referred to the original papers \cite{Dar:flux,MKV:PR}.

Using the above formulas we compute the shell-to-shell energy transfer
in MHD turbulence by summing up the energy transfer among the Fourier
modes. The energy transfer rates from $m$-th shell of field $X$
($u$ or $b$) to $n$-th shell of field $Y$ $(u$ or $b$) is\[
T_{nm}^{YX}=\sum_{\mathbf{k'}\in n}\sum_{\mathbf{p}\in m}\left\langle S^{YX}(\mathbf{k'|p|q})\right\rangle .\]
The $\mathbf{p}$-sum is over $m$-th shell, and the $\mathbf{k'}$-sum
is over $n$-th shell \cite{Lesi:book,MKV:PR}. Since $S^{YX}(\mathbf{k'}|\mathbf{p}|\mathbf{q})$
is a fluctuating quantity, we perform ensemble average to compute
the average shell-to-shell energy transfer. We compute the ensemble
average of $S$ using a standard field-theoretic technique, a technique
similar to EDQNM calculation \cite{Krai:65,Lesl:book,McCo:book}.
The calculation is quite standard, and it can found in McComb \cite{McCo:book}
or Verma \cite{MKV:PR}. 

The function $\left\langle S^{YX}(\mathbf{k'|p|q})\right\rangle $
depends on kinetic energy $(E^{u}(k))$, magnetic energy ($E^{b}(k))$,
cross helicity $(H_{c}(k))$, magnetic helicity $(H_{M}(k))$, kinetic
helicity $(H_{K}(k))$, and mean magnetic field. The total cross helicity,
magnetic helicity, and kinetic helicity are defined as $\mathbf{u}\cdot\mathbf{b}/2$,
$\mathbf{\mathbf{a}\cdot\mathbf{b}/2}$, and $\mathbf{u}\cdot\mathbf{\omega}/2$
respectively, where $\mathbf{a}$ and $\mathbf{\omega}$ are vector
potential and vorticity respectively. The spectrum of these quantities
are defined appropriately (refer to Verma \cite{MKV:PR} for details).
For simplification we compute $S$ for zero cross helicity and zero
magnetic field, and in the inertial range using Kolmogorov's energy
spectrum. To study the effects of kinetic and magnetic helicities,
we have split $S$ into helical and nonhelical components. The simplified
expression (given below) is a function of Alfvén ratio $(r_{A}=E^{b}(k)/E^{u}(k)$),
normalized magnetic helicity $(r_{M}=kH_{M}(k)/E^{b}(k))$, and normalized
kinetic helicity $(r_{K}=H_{K}(k)/(kE^{u}(k)))$. Please note that
we are working in three dimensions.

After lengthy algebra we obtain\begin{equation}
\frac{\left\langle S^{XY}(v,w)\right\rangle }{\Pi}=\left[\frac{(K^{u})^{3/2}(2\pi)^{2}}{k^{6}}\right]\left[F_{nonhelical}^{XY}(v,w)+F_{helical}^{XY}(v,w)\right].\label{eq:Savg}\end{equation}
where $K^{u}$ is Kolmogorov's constant for MHD turbulence, and some
of the $F_{nonhelical}^{YX}$ and $F_{helical}^{YX}$ are \begin{eqnarray}
F_{nonhelical}^{bb} & = & \frac{\frac{1}{r_{A}}t_{4}(v,w)(vw)^{-11/3}+\frac{1}{r_{A}}t_{8}(v,w)w^{-11/3}+\frac{1}{r_{A}^{2}}t_{10}(v,w)v^{-11/3)}}{\eta^{*}\left(1+v^{2/3}\right)+\nu^{*}w^{2/3}},\label{eq:F-nonhelical}\\
F_{nonhelical}^{ub} & =- & \frac{\frac{1}{r_{A}^{2}}t_{2}(v,w)(vw)^{-11/3}+\frac{1}{r_{A}}t_{7}(v,w)w^{-11/3}+\frac{1}{r_{A}}t_{11}(v,w)v^{-11/3)}}{\eta^{*}\left(v^{2/3}+w^{2/3}\right)+\nu^{*}},\\
F_{nonhelical}^{bu} & = & -\frac{\frac{1}{r_{A}}t_{3}(v,w)(vw)^{-11/3}+\frac{1}{r_{A}^{2}}t_{6}(v,w)w^{-11/3}+\frac{1}{r_{A}}t_{12}(v,w)v^{-11/3)}}{\eta^{*}\left(1+w^{2/3}\right)+\nu^{*}v^{2/3}},\\
F_{helical}^{bb} & = & \frac{\frac{r_{M}r_{K}}{r_{A}}t'_{4}(v,w)(vw)^{-11/3}+\frac{r_{M}r_{K}}{r_{A}}t'_{8}(v,w)w^{-11/3}+\frac{r_{M}r_{M}}{r_{A}^{2}}t'_{10}(v,w)v^{-11/3}}{\eta^{*}\left(1+v^{2/3}\right)+\nu^{*}w^{2/3}}.\label{eq:F-helical}\end{eqnarray}
Here$\nu^{*}$and $\eta^{*}$ are renormalized viscosity and resistivity
parameters. In this paper, we consider (a) nonhelical MHD ($r_{M}=r_{K}=0$
and different $r_{A}$s), and (b) helical MHD $(r_{A}=1,r_{K}=0.1,r_{M}=-0.1)$.
For nonhelical MHD with $r_{A}=0.5,1,2$, the constants $(K^{u},\nu^{*},\eta^{*})$
taken are $(0.55,2.1,0.5),$ $(0.75,1.0,0.69)$, $(1.0,0.64,0.77)$
respectively (see Verma \cite{MKV:MHD_RG,MKV:MHD_Flux} for the procedure
to compute these constants). For helical MHD, our choice of $r_{K}=0.1$
(small positive) and $r_{M}=-0.1$ (small negative) is one of the
typical values taken in numerical simulations, or observed in astrophysical
situations; for this case, the constants $K^{u}=0.78,\,\nu^{*}=1.0,\,\eta^{*}=0.69$
have been taken from Verma \cite{MKV:MHD_Helical}. 

The wavenumbers shells are binned logarithmically with the $n$-th
shell being $(k_{0}s^{n-1},k_{0}s^{n})$. Note that the parameter
$s$ is similar to the scale-disparity parameter of Zhou \cite{Zhou:Local}.
We nondimensionalize the equations using the transformation \cite{Lesl:book}\begin{equation}
k=\frac{a}{u};\,\,\,\,\, p=\frac{a}{u}v;\,\,\,\,\, q=\frac{a}{u}w,\end{equation}
 where $a=k_{0}s^{n-1}$. The resulting equation is\begin{eqnarray}
\frac{T_{nm}^{YX}}{\Pi} & = & K_{u}^{3/2}\frac{1}{2}\int_{s^{-1}}^{1}\frac{du}{u}\int_{us^{m-n}}^{us^{m-n+1}}dv\int_{|1-v|}^{1+v}dw\left(vw\right)\left[F_{nonhelical}^{YX}(v,w)+F_{helical}^{YX}(v,w)\right],\label{eqn:shell_final}\end{eqnarray}
 which is independent of $a$ in the inertial range. The independence
of $a$ implies self-similarity in the inertial range. From Eq. (\ref{eqn:shell_final})
we can draw the following inferences:

\begin{enumerate}
\item The shell-to-shell energy transfer rate is a function of $n-m$, that
is, $\Phi_{nm}=\Phi_{(n-i)(m-i)}$. Hence, the turbulent energy transfer
rates in the inertial range are all self-similar. Note that this property
holds only in the inertial range.
\item $T_{nm}^{ub}/\Pi=-T_{mn}^{bu}/\Pi$, or $b$-to-$u$ energy transfer
rates from shell $m$ to shell $n$ is equal and opposite to the $u$-to\-$b$
energy transfer rates from shell $n$ to shell $m$. Hence $T_{nm}^{bu}/\Pi$
can be obtained from $T_{mn}^{ub}/\Pi$ by inversion at the origin. 
\item The MHD energy fluxes are related to the shell-to-shell energy transfers
by the relationship\[
\Pi_{Y>}^{X<}=\sum_{n=m+1}^{\infty}(n-m)T_{nm}^{YX}.\]

\item Net energy gained by a $u$-shell from $u$-to-$u$ transfer is zero
because of self similarity. However, a $u$-shell can gain or lose
a net energy due to imbalance between $u$-to-$b$ and $b$-to-$u$
energy transfers. By definition, we can show that net energy gained
by an inertial $u$-shell is\begin{equation}
\sum_{n}\left(T_{nm}^{ub}-T_{nm}^{bu}\right)+T_{nn}^{ub}.\label{eq:Sum-Tub}\end{equation}
Similarly, net energy gained by a $b$-shell from $b$-to-$b$ transfer
is zero. However, net energy gained by an inertial $b$-shell due
to $u$-to-\textbf{$b$} and $b$-to-$u$ transfers is \begin{equation}
\sum_{n}\left(T_{nm}^{bu}-T_{nm}^{ub}\right)+T_{nn}^{bu}.\label{eq:Sum-Tbu}\end{equation}

\end{enumerate}
Now we compute the integrals of Eq. (\ref{eqn:shell_final}); we denote
the nonhelical and helical parts by $\left(T_{nm}^{YX}\right)_{nonhelical}$
and $\left(T_{nm}^{YX}\right)_{helical}$ respectively. Their properties
are described below.

\subsubsection{Nonhelical shell-to-shell energy transfer}

\begin{figure}
\includegraphics[%
  scale=0.9]{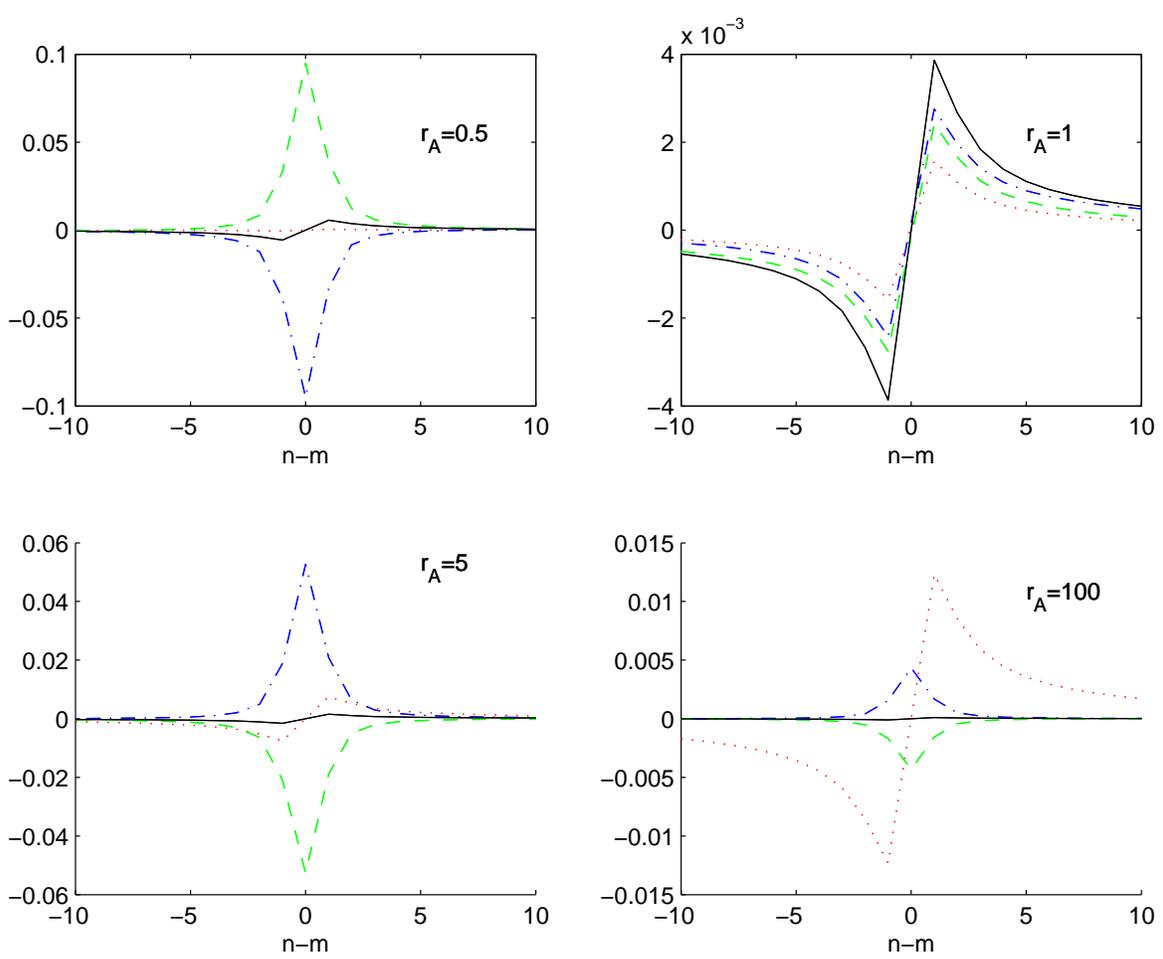}

\caption{\label{Fig:T-MHD-theory} The plots of shell-to-shell energy transfers
$\left(T_{nm}^{YX}\right)_{nonhelical}/\Pi$ vs. $n-m$ for zero helicities
$(\sigma_{c}=r_{K}=r_{M}=0)$ and Alfvén ratios $r_{A}=0.5,1,4,100$.
Here $s=2^{1/32}$. The $u$-to-$u$, $b$-to-$u$, $u$-to-$b$,
and $b$-to-$b$ are represented by dotted, dashed, chained, and solid
lines respectively. For $r_{A}=0.5$, the maxima of $\left(T_{nm}^{ub}\right)_{nonhelical}/\Pi$
and $\left(T_{nm}^{bu}\right)_{nonhelical}/\Pi$ are $\pm0.1$ respectively,
out of scale of the plot. The corresponding values for $r_{A}=5$
are $\mp0.053$.}
\end{figure}

We compute the nonhelical shell-to-shell energy transfer using $F_{nonhelical}^{YX}$.
We have chosen $s=2^{1/32}$. This study has been done for various
values of Alfvén ratios. Fig. \ref{Fig:T-MHD-theory} contains plots
of $\left(T_{nm}^{YX}/\Pi\right)_{nonhelical}$ vs. $n-m$ for four
typical values of $r_{A}=0.5,1,5,100$. The numbers on the plots represent
energy transfer rates from shell $m$ to shells $m+1,m+2,...$in the
right, and to shells $m-1,m-2,...$ in the left. For $r_{A}=0.5$,
the maxima of $b$-to-$u$ energy transfers $\left(T_{nm}^{ub}\right)_{nonhelical}/\Pi$
and $\left(T_{nm}^{bu}\right)_{nonhelical}/\Pi$ occurs at $m=n$,
and its values are approximately $\pm0.1$ respectively. The corresponding
values for $r_{A}=5$ are approximately $\mp0.053$. By observing
the plots we find the following interesting patterns:

\begin{enumerate}
\item The $u$-to-$u$ energy transfer rate from shell $m$ to shell $n$
$\left(T_{nm}^{uu}\right)_{nonhelical}/\Pi$ is positive for $n>m$,
and negative for $n<m$. Hence, a $u$-shell gains energy from smaller
wavenumber $u$-shells, and loses energy to higher wavenumber $u$-shells,
implying that the energy cascade is forward. Also, the absolute maximum
occurs for $n=m\pm1$, hence the energy transfer is local. For kinetic
energy dominated regime, $s=2^{1/2}$ yields $T_{nm}^{uu}/\Pi\approx35$\%,
similar to Kraichnan's Test Mean Field model (TFM) predictions \cite{Krai:71}. 
\item The $b$-to-$b$ energy transfer rate $T_{nm}^{bb}/\Pi$ is positive
for $n>m$, and negative for $n<m$, and maximum for $n=m\pm1.$ Hence
magnetic to magnetic energy transfer is forward and local. This result
is consistent with the forward magnetic-to-magnetic cascade $(\Pi_{b>}^{b<}>0)$
\cite{MKV:MHD_Flux,MKV:PR}.
\item For $r_{A}>1$ (kinetic energy dominated), kinetic to magnetic energy
transfer rate $\left(T_{nm}^{bu}\right)_{nonhelical}/\Pi$ is positive
\emph{}most of the shells. For $n-m<-30$ or so, the value is small
and negative. These transfers have been illustrated in Fig. \ref{Fig:ub-shell}(a).
Using Eq. (\ref{eq:Sum-Tbu}) we find that each $u$-shell loses a
net kinetic energy to $b$-shells, hence the turbulence is not steady.
This phenomena is seen for all $r_{A}>1$, and it could be one of
the processes responsible for dynamo action. For $s=2^{1/4}$, $\left(T_{nn}^{bu}\right)_{nonhelical}/\Pi\approx1.4$,
and the $\left(T_{nm}^{bu}\right)_{nonhelical}/\Pi$ is positive for
$n\ge m-1$ and negative otherwise.
\item For $r_{A}=0.5$ (magnetically dominated), magnetic to kinetic energy
transfer rate $\left(T_{nm}^{ub}\right)_{nonhelical}/\Pi$ is positive
for most of the shells (see Fig. \ref{Fig:T-MHD-theory}). For $n-m<-30$
or so, the value is small and negative. In addition, using Eq. (\ref{eq:Sum-Tub})
we find that each $b$-shell loses a net magnetic energy to $u$-shells,
hence the turbulence cannot be steady. This phenomena is seen for
all $r_{A}<1$. For $s=2^{1/4}$, $\left(T_{nm}^{ub}\right)_{nonhelical}/\Pi$
is positive for $n\ge m-1$ and negative otherwise.
\item The observations of (3) and (4) indicate that kinetic to magnetic
or the reverse energy transfer rate almost vanishes near $r_{A}=1$.
\emph{We believe that the evolution of MHD turbulence toward $r_{A}\approx1$
in both decaying and steady-state is due to the above reasons.} For
$r_{A}\ne1$, MHD turbulence is not steady. This result is similar
to Pouquet et al.'s prediction of equipartition of kinetic and magnetic
energy using EDQNM calculation \cite{Pouq:EDQNM}. An analogous result
was discovered by Stribling and Matthaeus \cite{Stri:Abso} in the
context of the Absolute Equilibrium Ensemble (AEE) theory. \\
The steady-state value of $r_{A}$ in numerical simulations (e.g.
Debliquy et al. \cite{Oliv:Simulation}, Dar et al. \cite{Dar:flux},
Alexakis et al. \cite{Mini:I}, Mininni et al. \cite{Mini:II}, and
Haugen et al. \cite{Bran:Nonhelical_simulation}) and solar wind (e.g.,
Matthaeus and Goldstein \cite{MattGold}) is around 0.5-0.6. The difference
is probably because the realistic flows have more interactions than
those discussed above, e.g., nonlocal coupling with forcing wavenumbers,
coherent dissipative structures \cite{MattLamk} etc. 
\item When $r_{A}$ is not close to 1 ($r_{A}\le0.5$ or $r_{A}>5$), $u$-to-$b$
shell-to-shell transfer involves many neighbouring shells (see Fig.
\ref{Fig:T-MHD-theory}). This observation implies that $u-$$b$
energy transfer is somewhat nonlocal as predicted by Pouquet et al.
\cite{Pouq:EDQNM}.
\item We compute energy fluxes using $T_{nm}^{YX}$, and find them to be
the same as that computed by Verma \cite{MKV:MHD_Flux,MKV:PR}. Hence
both the results, flux and shell-to-shell energy transfer rates, are
consistent with each other.
\end{enumerate}
Debliquy et al. \cite{Oliv:Simulation} performed a decaying MHD turbulence
simulations and computed the shell-to-shell energy transfer rates.
The kinetic and magnetic helicity was approximately zero. Debliquy
et al.'s results show that the shell-to-shell energy transfers are
forward and local. They also find that in the magnetically dominated
MHD, the energy transfer from the same shell is from magnetic field
to velocity field. Since the numerical simulations start with $r_{A}=1$,
the energy transfer rates for kinetic-energy dominated regime is not
known numerically. Our theoretical results are in general agreement
with the numerical results of Debliquy et al. As an example, the numerical
values of shell-to-shell energy transfer rates shown in Fig. 9 of
Debliquy et al. \cite{Oliv:Simulation} ($r_{A}\approx0.4$) is similar
to our theoretical results shown in Fig. \ref{Fig:T-MHD-theory}.
A major difference between theoretical and numerical values are for
$T^{ub}/\Pi$ where theoretical value for $n=m$ is larger than its
numerical counterpart. 

Debliquy et al. \cite{Oliv:Simulation} found that $T^{bu}$ changes
sign near $r_{A}\approx0.5$. In our theoretical calculation, the
change of sign takes place around $r_{A}\approx1$. These findings
are very encouraging, and they yield explanation why the asymptotic
state of Alfvén ratio is close to 0.5. The difference between theory
and numerical simulations may be due to the neglect of large-scale
non-Kolmogorov-like behaviour in theory, or due to the fact that Debliquy
et al.'s results are based on decaying simulations, while the theoretical
results assume steady-state. These issues need to be addressed.

Alexakis et al. \cite{Mini:I} and Mininni et al. \cite{Mini:II}
computed shell-to-shell energy transfers in forced MHD. In the inertial
range, the energy transfers are essentially local. However, the forcing
velocity-shell (at large length-scale) provides energy to the large-scale
magnetic field. The forcing velocity-shells also provide energy to
the inertial range shells by \emph{nonlocal} channel. Similar picture
was observed by Dar et al. \cite{Dar:flux} in forced 2D MHD turbulence.
Our theoretical results are consistent with the Alexakis et al.'s
results in the inertial range. Unfortunately, the present theoretical
calculation cannot predict the coupling with the forcing shell. 

After the above discussion on nonhelical MHD, we move to helical MHD.

\subsubsection{Helical Contributions}

\begin{figure}
\includegraphics{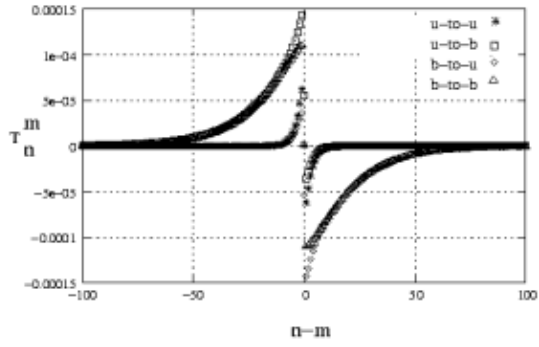}

\caption{\label{Fig:T-MHD-helical-theory}\emph{Helical contributions to}
shell-to-shell energy transfer\emph{s$\left(T_{nm}^{YX}\right)_{helical}/\Pi$}
vs. $n-m$ in helical MHD with $r_{A}=1,r_{K}=0.1,r_{M}=-0.1$ and
$\sigma_{c}=0$. Here $s=2^{1/4}$.}
\end{figure}

\begin{figure}
\includegraphics{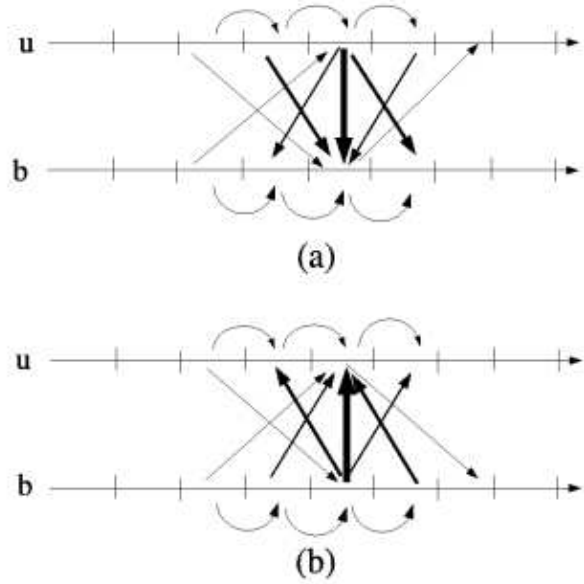}

\caption{\label{Fig:ub-shell} Schematic illustration of nonhelical shell-to-shell
energy transfers$T_{nm}^{YX}/\Pi$ in the inertial range for (a) kinetic-energy
dominated regime, and (b) magnetic-energy dominated regime. In (a)
the $u$-to-$b$ energy transfer rate $T_{nm}^{bu}/\Pi$ is positive
for most $n$, while in (b) the $b$-to-$u$ energy transfer rate
$T_{nm}^{ub}/\Pi$ is positive for most $n$. The $u$-to-$u$ energy
transfer rate $T_{nm}^{uu}$, and the $b$-to-$b$ energy transfer
rate $T_{nm}^{bb}$ are forward and local. }
\end{figure}

Now we present computation of $\left(T_{nm}^{YX}\right)_{helical}/\Pi$,
shell-to-shell energy transfer rates for helical MHD $(H_{M}\ne0,H_{K}\ne0)$
\cite{MKV:MHD_Helical}. To simplify the equation, we consider only
nonAlfvénic fluctuations ($\sigma_{c}=0$). We have chosen $r_{A}=1,r_{K}=0.1,r_{M}=-0.1$.
These values are one of the typical parameter values chosen in numerical
simulations. We take $s=2^{1/4}$ to get a increased value for $\left(T_{nm}^{YX}\right)_{helical}/\Pi$.
For the above choice of parameters, Kolmogorov's constant $K^{u}=0.78$
\cite{MKV:MHD_Helical}. In Fig. \ref{Fig:T-MHD-helical-theory} we
have plotted $\left(T_{nm}^{YX}\right)_{helical}/\Pi$ vs $n-m$.
Our results on helical shell-to-shell transfers are given below:

\begin{enumerate}
\item Comparison of Fig. \ref{Fig:T-MHD-helical-theory} with Fig. \ref{Fig:T-MHD-theory}
($r_{A}=1$) shows that helical energy transfers are order-of-magnitude
lower than the nonhelical ones for the parameters chosen here ($r_{A}=1,r_{K}=0.1,r_{M}=-0.1$).
For maximal helicity, the helical and nonhelical values become comparable.
\item All the helical contributions are negative for $n>m$, and positive
for $n<m$. Hence, helical transfers are from larger wavenumbers to
smaller wavenumbers. This is consistent with the inverse cascade of
energy due to helical contributions, as discussed by Pouquet et al.
\cite{Pouq:EDQNM,MKV:MHD_Helical}.
\item We find that the helical shell-to-shell energy transfer rate $\left(T_{nm}^{ub}\right)_{helical}$
and $\left(T_{nm}^{bb}\right)_{helical}$ is significantly positive
for $-50<n-m\le0$. This signals a nonlocal $b$-to-$b$ and $u$-to-$b$
inverse energy transfers. Hence, helicity induces nonlocal energy
transfer between $b$-to-$b$ and $u$-to-$b$ wavenumber shells.
This is in agreement with Pouquet et al.'s result \cite{Pouq:EDQNM}
that {}``residual helicity'' induces growth of large-scale magnetic
field by nonlocal interactions. 
\end{enumerate}
The theoretical findings listed above are consistent with Pouquet's
results based on EDQNM approximation \cite{Pouq:EDQNM} and flux calculations
of Brandenburg et al. \cite{Bran:Alpha}. Alexakis et al. \cite{Mini:I}
and Mininni et al. \cite{Mini:II} have computed shell-to-shell energy
transfer in helical MHD turbulence; the helicity does change the energy
transfer rates, however, in the absence of numerical value of normalized
kinetic and magnetic helicity $(r_{K},r_{M})$, we are not able to
compare our results with their numerical values.

In the next section we use our theoretical results and Debliquy et
al.'s \cite{Oliv:Simulation} numerical results to argue why the asymptotic
state of MHD flows have $r_{A}\approx0.4-0.6$.

\section{Connection with MHD Asymptotic State ($r_{A}\approx0.4-0.6$)}

The solar wind observations and numerical simulations show that the
asymptotic state in the MHD flows have $r_{A}\approx0.4-0.6$. We
find in our theoretical analysis that for $r_{A}>1$, there is a preferential
transfer of kinetic energy to magnetic energy; in fact, a $u$-shell
loses a net amount of kinetic energy to $b$-shell. For $r_{A}<1$,
the pattern of $u$-to-$b$ energy transfer is reversed, and there
is a a net transfer of energy from magnetic to kinetic. This preferential
energy transfer is minimum for $r_{A}\approx1$. Our theoretical calculation
however is based on an assumption of Kolmogorov's spectrum for the
energy, which is not valid for the smaller wavenumber modes. Here
we use Debliquy et al.'s \cite{Oliv:Simulation} decaying simulations
results for obtaining further insights into the large-scale energy
transfers. 

Debliquy et al.'s \cite{Oliv:Simulation} showed that the energy flux
from small-wavenumber $u-$sphere to small wavenumber $b$-sphere
is positive for $r_{A}>0.63$, and becomes negative for lower $r_{A}$.
The global (inclusive of all shells) $u$-to-$b$ energy transfer
changes sign at $r=0.4$. Hence, the smaller wavenumbers shells also
play an important role in energy transfers. As a result, the asymptotic
state of MHD turbulence has Alfvén ration close to 0.5. Thus our detailed
analysis of shell-to-shell energy transfer, and Debliquy et al.'s
global and flux analysis is able to explain qualitatively why MHD
turbulence evolves to asymptotic state with Alfvén ratio $r_{A}\approx0.4-0.6$.
Our theoretical predictions are consistent with the numerical results
of Dar et al. \cite{Dar:flux} whose asymptotic Alfvén ratio $r_{A}$
is approximately 0.5. Alexakis et al. \cite{Mini:I} and Mininni et
al. \cite{Mini:II} also find their asymptotic Alfvén ratio $r_{A}$
to be less than 1.

\section{Conclusions}

In this paper we have computed the shell-to-shell energy transfers
in MHD turbulence analytically. Our results provide theoretical explanation
for the recently computed shell-to-shell energy transfers using direct
numerical simulations. The contributions of nonhelical and helical
terms have been calculated separately. We find that the nonhelical
$u$-to-$u$ and $b$-to-$b$ shell-to-shell energy transfers are
local as in fluid turbulence, i. e., most of energy from a wavenumber
shell is transferred to the neighbouring shells. Comparatively, helical
$u$-to-$b$ and $b$-to-$u$ energy transfers involves distant shells
(nonlocal). 

We find that the helical shell-to-shell energy transfer is backward,
that is from larger wavenumbers to smaller wavenumbers. For $r_{K}=0.1,r_{M}=-0.1$,
one of the typical values observed in numerical simulations, the helical
shell-to-shell energy transfer is order-of-magnitude smaller than
the nonhelical ones. However, for maximal helicity, the helical shell-to-shell
energy transfer is comparable to the nonhelical ones. In the present
calculation, the parameters $r_{K}$ and $r_{M}$ have been chosen
to be constants. This is a gross assumption considering that magnetic
helicity and kinetic helicity have different signs at different scales.
Even then we obtain results which are consistent with recent numerical
results and earlier theories (Frisch et al. \cite{Fris:HM}). Hence,
the present calculation appears to capture some of the essential features
of helical MHD turbulence.

Our results show that the inertial-range shell-to-shell for $r_{A}>1$,
there is a preferential transfer of kinetic energy to magnetic energy;
the direction of energy transfer switches for $r_{A}<1$. Debliquy
et al.'s \cite{Oliv:Simulation} numerical simulations provide us
important clues for the energy transfers at smaller wavenumbers. Using
these results we can argue why the asymptotic state of MHD turbulence
evolves to Alfvén ratio of 0.5.

Our theoretical results are in general agreement with the numerical
results of Debliquy et al. \cite{Oliv:Simulation}, Dar et al. \cite{Dar:flux},
Alexakis et al. \cite{Mini:I} and Mininni et al. \cite{Mini:II},
who observe local energy transfer in the inertial range. In forced
MHD turbulence with large-scale velocity forcing, a significant energy
transfer to large-scale magnetic field and nonlocal energy transfer
to inertial-range magnetic field were found. Since the nonlinear energy
transfers involves only the velocity and magnetic field variables,
we expect that the features of energy exchange in the inertial range
should remain approximately the same in decaying and forced MHD. However,
the forcing at large-scale velocity shell would affect the large-scale
magnetic field, and could also induce nonlocal interations. A theoretical
model with forcing will be useful to understand forced MHD turbulence. 

Detailed pictures of energy transfer studied here is useful for understanding
various physical process, for example, dynamo mechanism to generate
magnetic field in astrophysical objects. These results are also useful
in modelling MHD flows and in simulations. For example, we need to
model backscatter and forward energy transfer for large-eddy simulations.
Refer to Debliquy et al. \cite{Oliv:Simulation} for discussion on
backscatter and forward energy transfer in Fourier as well as real
space.

In EDQNM analysis, the wavenumber shells are logarithmically binned.
In the present analysis of MHD turbulence, and many papers on fluid
turbulence show local energy transfers among wavenumber shells for
nonhelical MHD. Hence, the local energy transfer assumptions made
in EDQNM analysis is valid at least for nonlocal MHD \cite{Zhou:RMP}.
The energy transfers are somewhat more complex for helical MHD. 

Pouquet et al. \cite{Pouq:EDQNM} performed extensive EDQNM analysis
of MHD turbulence and showed that both local and nonlocal interactions
exist in MHD turbulence. The local interactions cause the energy cascade,
while the nonlocal ones cause equipartition of kinetic and magnetic
energies. Pouquet et al. argued for inverse cascade of magnetic energy
from the competition between helicity and Alfvén effect. Our analytic
calculation also predicts inverse magnetic-energy cascade due to helicity.
Our calculation shows that near equipartition of magnetic and kinetic
energy is due to complex process involving inertial-range shell-to-shell
interactions and small wavenumber shells. This picture is some what
different than that of Pouquet et al. 

To conclude, the shell-to-shell energy transfer rates provide important
insights into inertial-range energy exchange processes in MHD turbulence.

\begin{acknowledgments}
MKV gratefully acknowledges useful comments and suggestions by Diego
Donzis, Daniele Carati, and Olivier Debliquy. Part of the work was
supported by a project from Department of Science and Technology,
India.
\end{acknowledgments}

\bibliographystyle{apsrev}


\begin{center}{\Large Figure Captions}\end{center}{\Large \par}

\textbf{Figure 1:} The plots of shell-to-shell energy transfers $\left(T_{nm}^{YX}\right)_{nonhelical}/\Pi$
vs. $n-m$ for zero helicities $(\sigma_{c}=r_{K}=r_{M}=0)$ and Alfvén
ratios $r_{A}=0.5,1,4,100$. Here $s=2^{1/32}$. The $u$-to-$u$,
$b$-to-$u$, $u$-to-$b$, and $b$-to-$b$ are represented by dotted,
dashed, chained, and solid lines respectively. For $r_{A}=0.5$, the
maxima of $\left(T_{nm}^{ub}\right)_{nonhelical}/\Pi$ and $\left(T_{nm}^{bu}\right)_{nonhelical}/\Pi$
are $\pm0.1$ respectively, out of scale of the plot. The corresponding
values for $r_{A}=5$ are $\mp0.053$.

\vspace{1cm}

\textbf{Figure 2:}\emph{Helical contributions to} shell-to-shell energy
transfer\emph{s$\left(T_{nm}^{YX}\right)_{helical}/\Pi$} vs. $n-m$
in helical MHD with $r_{A}=1,r_{K}=0.1,r_{M}=-0.1$ and $\sigma_{c}=0$.
Here $s=2^{1/4}$.

\vspace{1cm}

\textbf{Figure 3:} Schematic illustration of nonhelical shell-to-shell
energy transfers$T_{nm}^{YX}/\Pi$ in the inertial range for (a) kinetic-energy
dominated regime, and (b) magnetic-energy dominated regime. In (a)
the $u$-to-$b$ energy transfer rate $T_{nm}^{bu}/\Pi$ is positive
for $n\ge m-1$, and negative otherwise, while in (b) the $b$-to-$u$
energy transfer rate $T_{nm}^{ub}/\Pi$ is positive for $n\ge m-1$,
and negative otherwise. The $u$-to-$u$ energy transfer rate $T_{nm}^{uu}$,
and the $b$-to-$b$ energy transfer rate $T_{nm}^{bb}$ are forward
and local. 

\newpage

\begin{center}FIGURES\end{center}

\newpage

\newpage

\end{document}